\documentstyle[preprint,aps,epsf]{revtex}

\begin{document}

\title{
Hybrid Decays}

\author{Eric S. Swanson$^{1,2}$ and Adam P. Szczepaniak$^{1}$}

\address{$^1$\ 
   Department of Physics,
   North Carolina State University,
   Raleigh, North Carolina 27695-8202\\
	 $^2$\ 
   Jefferson Laboratory, 
   12000 Jefferson Avenue, 
   Newport News, VA 23606} 

\maketitle

\begin{abstract}
The heavy quark expansion of Quantum Chromodynamics and the strong coupling 
flux tube picture of
nonperturbative glue are employed to develop the phenomenology of hybrid meson 
decays. 
The decay mechanism explicitly couples gluonic degrees of freedom to the pair
produced quarks and hence does {\it not} obey the well known, but model-dependent,
selection rule which states that hybrids do not decay to pairs of $L=0$ mesons.
However, the nonperturbative nature of gluonic excitations in the flux tube 
picture leads to a new selection rule: light hybrids do not 
decay to pairs of identical mesons. New features of the model are highlighted 
and partial widths are presented for several low lying hybrid states.

\end{abstract}
\date{April, 1997}
\pacs{}
\narrowtext

\section{Introduction} 

Although explicit nonperturbative gluonic degrees of freedom
have been anticipated for many years, no clear experimental evidence for
their existence has emerged.
The search for nonperturbative glue, in particular as manifested in hybrid 
mesons, would be greatly
facilitated by a rudimentary knowledge of the hybrid spectrum and decay
characteristics. Although it appears that lattice estimates of light quenched
hybrid masses are forthcoming\cite{UKQCD}, hadronic decays remain difficult
to calculate on the lattice. Thus one is forced to rely on model estimates
of the couplings of hybrids to ordinary mesons. Historically, there have 
been two 
approaches to such estimates. The first assumes that hybrids are 
predominantly quark-antiquark states with an additional constituent 
gluon\cite{HM} and that decays proceed via constituent gluon 
dissociation\cite{T}. The second assumes that hybrids are predominantly
quark-antiquark states moving on an adiabatic surface generated by an excited 
``flux tube" configuration of glue\cite{IP}. Decays then proceed by a 
phenomenological pair production mechanism (the ``$^3P_0$ model") coupled 
with a flux tube overlap\cite{IKP}.

The authors of Refs. \cite{T} noted the existence of a selection rule: TE
hybrids do not decay to identical S-wave mesons. A similar observation was
made by Isgur, Kokoski, and Paton\cite{IKP} (hereafter referred to as IKP),
{\em no} hybrids decay to identical S-wave mesons\footnote{Note that
the analogue of TE hybrids does not exist in the flux tube model.}. Although this
assertion has achieved the status of dogma in hadron phenomenology, we stress
that it is model-dependent. In particular, in the work of IKP the selection rule
follows from the assumption that the quark pair production operator is completely
decorrelated with the gluonic modes in the hybrid (as it is in the
successful $^3P_0$ model\cite{TP0}). This implies that the hybrid flux tube degrees 
of freedom must annihilate the final state mesonic flux tubes. The result is
a spatial flux tube overlap function which defines the region in which the
pair creation may occur, has a node along the hybrid quark-antiquark
axis, and leads to the above mentioned selection rule. However, it is entirely 
feasible for
the hybrid flux tube to be correlated with the pair produced
quarks. In this case the argument of IKP breaks down and the selection rule
need not exist. This is precisely what happens in the model we develop here.

Our approach is based on the successful description of the Dirac structure
of confinement employed in Ref.\cite{SS}. The idea is to use the heavy quark
expansion of the Coulomb gauge QCD Hamiltonian to identify relevant operators. 
The gluonic portion of these are then evaluated using a slightly
extended version of the flux tube model of Isgur and Paton\cite{IP}.
In the heavy quark limit the quark and antiquark decouple at leading order.
We label these $\chi$ and $h$ 
where $h^\dagger$ creates a quark and $\chi$ creates an antiquark. 
In this approach the interaction contains the kinetic energy term: $H_{int} = -\int h^\dagger 
\bbox{\sigma} \cdot {\bf D} \chi + {\rm H.c.}$, where 
${\bf D} = i \nabla + g {\bf A}$. It should be noted that this interaction 
only contains terms which produce or annihilate a $Q\bar Q$ pair (i.e., gluon
production or annihilation by a through-going quark line are not present). This
implies that contributions to the decay which are higher order in the S-matrix
expansion are also higher order in $1/M_Q$. Our model is therefore rather simple:
calculate the decay of a (flux tube model) hybrid into two ordinary mesons
to first order in $H_{int}$. The essential new feature is that the gluon field
operator should be expressed in terms of the nonperturbative phonon modes
of the flux tube model -- as opposed to the traditional expansion in 
plane wave gluons.

\section{Flux Tube Model of Hybrid Decays}

As discussed above, we shall employ the flux tube model of Isgur and Paton \cite{IP} to construct 
the hybrid states and  the effective decay operator.
The model is extracted from the strong coupling limit of the QCD lattice 
Hamiltonian. The Hamiltonian is first split into blocks of distinct
``topologies" (in reference to possible gauge invariant flux tube 
configurations)  and then adiabatic and  small oscillation approximations
of the flux tube dynamics are made to arrive at an N-body discrete string-like 
model Hamiltonian
for gluonic degrees of freedom. This is meant to be operative at intermediate 
scales $a \sim b^{-1/2}$ where the strong coupling is of order unity.
The lattice spacing is denoted by $a$ and there are $N$ 
``beads'' (or links) evenly spaced between the $Q \bar Q$ pair. Diagonalizing 
the Flux Tube Hamiltonian yields phonons, $\alpha^a_{m,\lambda}$, which 
are labelled by their color ($a$), mode number ($m$), and polarization ($\lambda$)
which is transverse to the $Q\bar Q$ axis. A hybrid may be built
of $n_{m\lambda}$ phonons in the m'th mode with polarization $\lambda = \pm$.
In particular, hybrid states with a single phonon excitation are constructed as

\begin{equation}
\vert H \rangle \sim   \int d {\bf r} 
\, \varphi_H(r) \, \chi^{PC}_{\Lambda,\Lambda'}\, 
D^{L_H*}_{M_L,\Lambda}(\phi,\theta,-\phi) \, T^a_{ij}\, b_i^\dagger({\bf r}/2) 
d_j^\dagger(-{\bf r}/2) \alpha^{a\dagger}_{m,\Lambda'} \vert 0 \rangle.
\end{equation}

\noindent
Spin and flavor indices have been suppressed and color indices are explicit.
The factor $\chi^{PC}_{\Lambda,\Lambda'}$ in the hybrid wavefunction projects
onto states of good parity and 
charge conjugation. The quantum numbers of these 
states\footnote{These expressions differ from Isgur and 
Paton\cite{IP} because we have adopted the standard definitions for the polarization
vectors and the Wigner
rotation matrix, following the Jacob-Wick conventions.} are given by
$P = \eta_{PC} (-)^{L_H + 1}$ and $C = \eta_{PC} (-)^{L_H + S_H + N}$ where
$\eta_{PC} = \chi^{PC}_{-1,-1} = \pm 1$ and 
$N = \sum_m m(n_{m+} + n_{m-})$. We shall consider low-lying hybrids only
 so that $m=1$ in what follows.

The definition of the hybrid state makes it clear that decays proceed via
the vector potential portion of the covariant derivative (recall that we work
at leading order in $1/M_Q$). Thus we proceed
by writing the vector potential in terms of phonon operators such that the
%
%
expressions for the electric and magnetic
fields employed in Ref. \cite{SS} are recovered in the continuum limit. 
This yields the following effective decay operator

\begin{equation}
H_{int} = {i g a^2 \over \sqrt{\pi}}  \sum_{m,\lambda} \int_0^1 d \xi 
\cos(\pi \xi)  T^a_{ij}\, h^\dagger_i(\xi {\bf r}_{Q\bar Q}) 
{\bbox\sigma}\cdot \hat{\bf e}_{\lambda}(\hat{\bf r}_{Q\bar Q}) \left(\alpha^a_{m \lambda} 
 - \alpha^{a \dagger}_{m \lambda} \right) \chi_j(\xi {\bf r}_{Q\bar Q}),
\end{equation}

\noindent 
where the $\hat{\bf e}(\hat{\bf r})$ are polarization vectors orthogonal to $\hat {\bf r}$.
The integral is defined along the $Q\bar Q$
axis only. The integration over the transverse directions yields the 
factor $a^2$ which may be interpreted as the transverse size of the flux tube. 
Note that the phonon operators represent gluonic excitations which are 
perpendicular to the $Q\bar Q$
axis. Although this appears problematical in traditional perturbation theory,
it is required here because, in the adiabatic limit, the gluonic field 
configuration must be defined in terms of the quark 
configuration and therefore the field expansion of the vector potential
depends on the quark state under consideration. 

The decay amplitude for a hybrid $H$ into mesons $A$ and $B$ is then given 
by:

\begin{eqnarray}
\langle H \vert H_{int}\vert AB\rangle &=& i {g a^2\over \sqrt{\pi}} {2\over 3}
  \int_0^1 d \xi \int d{\bf r} \cos(\pi \xi) \sqrt{2L_H+1\over 4 \pi} 
  {\rm e}^{i {\bf p}\cdot {\bf r}\over 2} \varphi_H(r) \varphi^*_A(\xi{\bf r}) 
  \varphi^*_B((1-\xi){\bf r}) \cdot \nonumber \\
  && \left[ {\cal D}^{L_H*}_{M_L \Lambda}(\phi,\theta,
  -\phi) \chi_{\Lambda,\lambda}^{PC} \hat {\bf e}_{\lambda}(\hat {\bf r}) \cdot
  \langle \bbox{\sigma} \rangle \right]
\end{eqnarray}

\noindent
where $\langle \bbox{\sigma}\rangle$ is the matrix element of the Pauli matrices 
between quark spin wavefunctions.
This  amplitude should be multiplied by
the appropriate flavor overlap and symmetry factor.

We note the following general properties of the decay amplitude. The operator 
is nonzero only along the hybrid $Q\bar Q$ axis -- as follows from the structure
of the interaction Hamiltonian. Thus $q\bar q$ creation occurs on a line joining
the original $Q\bar Q$ quarks, smeared over the transverse size of the flux
tube. This is in contrast to the model of IKP which has a node along the $Q\bar Q$
axis. Furthermore the spin operator contracts with the flux tube
phonon polarization vector, which is absent in the IKP model.
Finally, the decay amplitude vanishes when the final mesons are identical for
any single-phonon hybrid in an odd mode due to the nodal structure in 
the vector potential. Thus one obtains the selection rule:
low-lying hybrids do not decay to identical mesons. This subsumes the 
selection rule of IKP so that none of their qualitative conclusions are 
changed. However we also predict, for example, that hybrids do not decay
to pairs of identical P-wave mesons.

\section{Applications}

All strong hadronic decay calculations depend heavily on the hadronic 
wavefunctions employed. This is especially true here when final states
with similar spatial wavefunctions are considered. We have taken the simple
approach of adopting simple harmonic oscillator wavefunctions with width 
parameters  
chosen to minimize the eigenenergies of a constituent 
quark model Hamiltonian\cite{S}. Several of the predicted widths were compared to 
calculations with exact wavefunctions -- the results differed by less than
10\%, leading us to expect that the widths presented below are reasonably
reliable. Nevertheless dramatically different wavefunction width parameters
 have been
used in the literature (cf., the third reference of Ref. \cite{TP0}) so
that the reader is cautioned that these results are qualitative (pending
more detailed analysis\cite{PSS}). 

In the following, the normalization is fixed to yield a total width for the
$\hat \pi(1.8)$ of 220 MeV, close to the
experimental value of 212 $\pm$ 37 MeV for the $\pi(1.8)$. This gives $g
a^2 = 2.56$ GeV$^{-2}$. Notice that this is consistent with the strong 
coupling relationship which relates the string tension to these parameters, 
$b = g^2 C_F/2 a^2$, if one assumes $g \sim 1$.
Table I presents the decay widths for 
isoscalar, isovector, and $s\bar s$ hybrids with $J^{PC} = 1^{-+}$, $0^{-+}$,
and $1^{--}$.

\begin{table}
\caption{Dominant Hybrid Decay Modes.}
\begin{tabular}{llccl}
 state & $J^{PC}$ & mode & partial waves & partial widths (MeV) \\
\tableline
I=1 $\ $ $\hat x(1.9)$ & $1^{-+}$ & $\pi b_1$       & S,D     & 87, 4 \\
                    &         & $K \bar K_1(1.4) {\phantom x}^{\rm a}$\tablenotetext[1]{A mixing angle of
                    34 degrees has been used for the $K_1(1.27)$ and $K_1(1.4)$.}   & S,D     & 24, 0 \\
                    &         & $\pi f_1$           & S,D     & 21, 0 \\
                    &         & $\pi\rho$           & P       & 11 \\
                    &         & $\eta a_1$          & S,D     & 9, 0 \\
                    &         & $K \bar K_1(1.27) {\phantom x}^{\rm a}$  & S,D     & 8, 0 \\
                    &         & $K\bar K^*$         & P       & 2 \\
&&&&\\
I=0 $\ $ $\hat y(1.9)$ & $1^{-+}$ & $\pi a_1$       & S,D      & 64, 1 \\
                    &         & $K \bar K_1(1.4)$  &  S,D      & 24, 0 \\
                    &         & $\eta f_1$          & S,D      & 8, 0 \\
                    &         & $K \bar K_1(1.27)$  & S,D      & 7, 0\\
                    &         & $K \bar K^*$        & P        & 2 \\
&&&&\\
$s\bar s$ $\ $ $\hat z(2.1)$ & $1^{-+}$ & $K \bar K_1(1.4)$& S,D   & 99, 0 \\
                    &         & $K \bar K_1(1.27)$   &  S,D    & 13, 0 \\
                    &         & $K \bar K^*$        &  P     & 7 \\
\tableline
I=1 $\ $ $\hat \rho (1.9)$ & $1^{--}$ & $\pi a_1$  &  S,D     & 87, 1 \\
                    &         & $K \bar K_1(1.27)$  &  S,D    & 13, 0\\
                    &         & $\pi\omega$         &  P      & 10\\
                    &         & $K \bar K_1(1.4)$  &  S,D     & 7, 0\\
                    &         & $\eta \rho$         &   P     & 5 \\
                    &         & $\eta' \rho$        &   P     & 1 \\
                    &         & $\pi a_2 $          &   D     & 1 \\
&&&&\\
+
I=0 $\ $ $\hat \omega(1.9)$ & $1^{--}$ & $\pi \rho$ & P  & 33 \\
                    &         & $K \bar K_1(1.27)$  & S,D  & 13, 0  \\
                     &         & $K \bar K_1(1.4)$  & S,D  & 7, 0  \\
                    &         & $\eta \omega$       & P   & 5\\
                    &         & $\eta' \omega$      & P & 1 \\
&&&&\\
$s\bar s$ $\ $ $\hat \phi(2.1)$ & $1^{--}$ & $K \bar K_1(1.4)$& S,D  & 53, 0 \\
                    &         & $K \bar K_1(1.27)$  &   S,D    & 23, 0 \\
                    &         & $\eta \phi$         &   P    & 10 \\
                    &         & $\eta' \phi$        &   P    & 3\\
                    &         & $ K_2^* \bar K$ &  D & 1 \\
\tableline
I=1 $\ $ $\hat \pi(1.8)$ & $0^{-+}$ & $ \bar{K} (K\pi)_{K_0^*} {\phantom x}^{b}$\tablenotetext[2]{The width of
                    the $K_0^*$ has been incorporated by convoluting with the
                    appropriate Breit-Wigner formula.} &     S   & 85 \\
                   &         & $\pi f_0(1.3) $     &    S    & 73 \\
                   &         & $\pi\rho$           &     P   & 38 \\
                   &         & $\eta a_0(1.3)$     &     S   & 14 \\
                   &         & $K \bar K^*$        &     P   & 6 \\
                   &         & $\pi\rho(1.465)$    &     P   & 4 \\
                   &         & $\pi f_2 $          &     D   & 1 \\
&&&&\\
I=0 $\ $ $\hat \eta(1.8) {\phantom{x}}^{\rm c}$\tablenotetext[3]{The flavor structure
of the $\hat \eta$ and $\hat \eta'$ are taken to be $(u\bar u + d \bar d)/\sqrt{2}$
and $s\bar s$ respectively.} & $0^{-+}$ & $\pi a_0(1.3)$   &    S     & 217 \\
                    &         & $\bar{K} (K\pi)_{K_0^*}$   &     S   & 85 \\
                    &         & $\eta f_0(1.3) $    &    S     & 14 \\
                    &         & $K \bar K^*$        &     P   & 6  \\
                    &         & $\pi a_2$     &   D      & 2 \\
&&&&\\
$s\bar s$ $\ $ $\hat \eta'(2.0) {\phantom{x}}^{\rm c}$ & $0^{-+}$ & $K  \bar K_0^*$&  S & 293 \\
                    &         & $K \bar K^*$        &     P   & 26  \\
                    &         & $K \bar K_2^*$      &    D     & 12 \\
                   
\end{tabular}
\label{tableI}
\end{table}

\section{Conclusions}

We have used an adiabatic approximation to hybrid structure to study the decay
of gluonic hybrid mesons.
Hybrids may be considered as quarks moving
on an adiabatic surface generated by a flux tube of nonperturbative glue 
in an excited state. In the heavy quark limit of QCD the decay operator 
is given by the transverse gluon component of the  
covariant derivative.
In keeping with the spirit of the flux tube model of Isgur and Paton, we 
chose to evaluate the gluonic portion of the decay amplitude in terms of 
phonon degrees of freedom. The resulting expression differs significantly 
from that of Ref.~\cite{IKP} --
but surprisingly gives similar predictions for the total widths of hybrids.
The main difference seems to be in substantially smaller D-wave amplitudes 
predicted here (this
is reminiscent of the difference between the $^3P_0$ model and the
$^3S_1$ model; see the last paper of Ref. \cite{TP0}). Qualitatively, the
calculation presented here obeys the well-known spin selection rule, spin zero
states do not decay to pairs of spin zero states, and a new rule: low lying
hybrids do not decay to pairs of identical mesons. Decays to mesons with 
similar spatial wavefunctions are also predicted to be suppressed.

Several candidate hybrid states exist. One of these is the $\pi(1800)$ 
seen in $\pi f_0(980)$, $\pi f_0(1300)$, and $K(K\pi)_S$\cite{VES} decay channels.
 Significantly,
the channels $\pi\rho$ and $KK^*$ are suppressed. Although a 3S quarkonium state
is expected near 1.8 GeV, the decay characteristics support the identification
of this state as a hybrid (see the discussion in Ref. \cite{BCPS} for a 
similar analysis). In particular the $\pi_{3S}$ is expected to decay strongly to
$\omega\rho$ while this mode is practically zero in our approach. The 
strength seen in
$\pi f_0(1300)$ is in accord with Table I and is again in conflict with a 
$q\bar q$
identification of the $\pi(1800)$. The interpretation of the $\pi f_0(980)$
mode is complicated by the large coupling of the $f_0(980)$ to the $K\bar K$ 
channel. 
Indeed, it is likely that the $f_0(980)$ is a $K\bar K$ bound state stablized by the 
$f_0(1300)$\cite{WI}. In view of this, the large $\pi f_0(980)$ mode may be 
due to strong final state interactions in $K \bar K_0^* \rightarrow K(\bar K\pi)$.
A search for isopartner states $\hat \eta$ and $\hat \eta'$ which are broad 
(roughly 330 MeV) and which do not couple to $\rho\rho$ or $\omega
\omega$ will be very instructive. 

Strong evidence for hybrids would be obtained if the  $1^{-+}$ seen at 1.9 GeV 
at BNL\cite{BNL} is confirmed since these quantum numbers are exotic. This state
has been seen in $\pi b_1$ and $\pi f_1$ in accordance with our expectations;
however, we also expect a reasonably large $K\bar K_1(1.4)$ mode and suggest 
that this should be 
investigated (although this might be complicated by strong $K\bar K$ final state
interactions). Finally, we consider the prospect for vector hybrid production at
TJNAF. If one assumes $\hat \rho$ and $\hat \omega$ production by vector meson
dominance followed by $t$-channel $\pi$ exchange then isospin symmetry implies
that the ratio of cross sections is $\sigma(\hat\omega)/\sigma(\hat\rho) = 27$.
Thus it would be expedient to search for the isoscalar vector hybrid. 
This is an interesting state because all of the normally large S+P channels are
excluded by quantum numbers and it is therefore very narrow, with a 
predicted width of roughly 60 MeV. The predominant decay mode is $\pi\rho$; we
therefore suggest 
that the search for the $\hat \omega$ be conducted in the 
$\pi^0\pi^+\pi^-$ final state.

\acknowledgements 
We are grateful to Ted Barnes, Suh Urk Chung, and Alex Dzierba for discussions 
on the phenomenology of hybrid states.
ES acknowledges the financial support of the DOE under grant DE-FG02-96ER40944.


\begin{references}

\bibitem{UKQCD} P. Lacock, C. Michael, P. Boyle, and P. Rowland, hep-lat/9611011.

\bibitem{HM} D. Horn and J. Mandula, Phys. Rev. {\bf D17}, 898 (1978); P. Hasenfratz,
R.R. Horgan, J. Kuti, and J.M. Richard, Phys. Lett. {\bf 95B}, 299 (1981); T. Barnes,
F.E. Close, and F. De Viron, Nucl. Phys. {\bf B224}, 241 (1983); M. Chanowitz
and S. Sharpe, Nucl. Phys. {\bf B222}, 211 (1983).

\bibitem{T} M. Tanimoto, Phys. Lett. {\bf 116B}, 198 (1982);
A. Le Yaouanc, L. Oliver, O. Pene, J.-C. Raynal, and S. Ono,
Z. Phys. C {\bf 28}, 309 (1985); F. Iddir, A. Le Yaouanc, L. Oliver, O. Pene,
and J.-C. Raynal, Phys. Lett. {\bf 207B}, 325 (1988).

\bibitem{IP} N. Isgur and J. Paton, Phys. Rev. {\bf D31}, 2910 (1985).

\bibitem{IKP} N. Isgur, R. Kokoski, and J. Paton, Phys. Rev. Lett {\bf 54}, 869
(1985); F.E. Close and P. Page, Nucl. Phys. {\bf B443}, 233 (1995).


\bibitem{TP0} L. Micu, Nucl. Phys. {\bf B10}, 521 (1969); R. Carlitz
and M. Kislinger, Phys. Rev. D {\bf2}, 336 (1970); R. Kokoski and N. Isgur, 
Phys. Rev. D {\bf35}, 907 (1987); A. Le Yaouanc, L. Oliver, O. Pene, and J.-C.
Raynal, Phys. Rev. D {\bf8}, 2233 (1973); Phys. Lett. {\bf 71 B}, 397
(1977); {\it ibid} {\bf 72 B}, 57 (1977); P. Geiger and E.S. Swanson, 
Phys. Rev. {\bf D50}, 6855 (1994). 


\bibitem{SS} A.P. Szczepaniak and E.S. Swanson, Phys. Rev. {\bf D55}, 3987 (1997).


\bibitem{S} E.S. Swanson, Ann. Phys. (NY) {\bf 220}, 73 (1992).

\bibitem{PSS} P. Page, E.S. Swanson, and A.P. Szczepaniak, in progress.

\bibitem{VES} D.V. Amelin {\it et al.}, Phys. Lett. {\bf B356}, 
595 (1995); I.A. Kachaev {\it et al.}, Phys. At. Nucl {\bf 57}, 1536 (1994); 
G. Bellini {\it et al.}, Phys. Rev. Lett. {\bf 48}, 1697 (1982).

\bibitem{BCPS} T. Barnes, F.E. Close, P. Page, and E.S. Swanson, Phys. Rev.
{\bf D55}, 4157 (1997).

\bibitem{WI} J. Weinstein and N. Isgur, Phys. Rev. {\bf D41}, 2236 (1990).

\bibitem{BNL} J.H. Lee {\it et al.}, Phys. Lett. {\bf B323}, 227 (1994).


\end{references}
\end{document}